# Light bending and X-ray echoes from behind a supermassive black hole

D.R. Wilkins[1]*, L.C. Gallo[2], E. Costantini[3,4], W.N. Brandt[5,6,7] and R.D. Blandford[1]

[1]Kavli Institute for Particle Astrophysics and Cosmology, Stanford University, 452 Lomita Mall, Stanford, CA 94305, USA
[2]Department of Astronomy & Physics, Saint Mary's University, Halifax, NS. B3H 3C3, Canada
[3]SRON, Netherlands Institute for Space Research, Sorbonnelaan 2, 3584 CA Utrecht, The Netherlands
[4]Anton Pannekoek Institute for Astronomy, University of Amsterdam, Science Park 904, 1098 XH Amsterdam, The Netherlands
[5]Department of Astronomy and Astrophysics, 525 Davey Lab, The Pennsylvania State University, University Park, PA 16802, USA
[6]Institute for Gravitation and the Cosmos, The Pennsylvania State University, University Park, PA 16802, USA
[7]Department of Physics, 104 Davey Lab, The Pennsylvania State University, University Park, PA 16802, USA
*Corresponding author. E-mail: dan.wilkins@stanford.edu

**The innermost regions of accretion disks around black holes are strongly irradiated by X-rays that are emitted from a highly variable, compact corona, in the immediate vicinity of the black hole [1, 2, 3]. The X-rays that are seen reflected from the disk [4] and the time delays, as variations in the X-ray emission echo or 'reverberate' off the disk [5, 6] provide a view of the environment just outside the event horizon. I Zwicky 1 (I Zw 1), is a nearby narrow line Seyfert 1 galaxy [7, 8]. Previous studies of the reverberation of X-rays from its accretion disk revealed that the corona is composed of two components; an extended, slowly varying component over the surface of the inner accretion disk, and a collimated core, with luminosity fluctuations propagating upwards from its base, which dominates the more rapid variability [9, 10]. Here we report observations of X-ray flares emitted from around the supermassive black hole in I Zw 1. X-ray reflection from the accretion disk is detected through a relativistically broadened iron K line and Compton hump in the X-ray emission spectrum. Analysis of the X-ray flares reveals short flashes of photons consistent with the re-emergence of emission from behind the black hole. The energy shifts of these photons identify their origins from different parts of the disk [11, 12]. These are photons that reverberate off the far side of the disk and bent around the black hole and magnified by the strong gravitational field. Observing photons bent around the black hole confirms a key prediction of General Relativity.**

I Zw 1 was observed simultaneously by the X-ray telescopes *NuSTAR* [13] and *XMM-Newton* [14], between 2020 January 11 and 2020 January 16. *NuSTAR* observed I Zw 1 in the hard X-ray band, detecting the emission between 3 and 50 keV, continuously over 5.3 days. *XMM-Newton* observed I Zw 1 over the 0.3-10 keV energy range, during two periods of 76 and 69 ks (Figure 1). About 150 ks from the start of the observations, flaring was observed in the X-ray emission. Two flares, each lasting 10 ks, were observed. The X-ray count rate peaked at 2.5 times the mean level before the flare.

The 3-50 keV X-ray spectrum is well described by an emission model consisting of the directly observed continuum from the corona and the reflection from the accretion disk around a rapidly spinning black hole with spin parameter $a > 0.75\, GM/c^2$, where $G$ is the gravitational constant, $M$ is the black hole mass and $c$ is the speed of light (Figure 2, Extended Data Figure 1).

The relativistically broadened iron K fluorescence line is seen around 6 keV in addition to the reflected Compton hump above 10 keV [15]. Line photons are emitted at 6.4 keV in the rest frame of the accretion disk but are observed shifted in energy by Doppler shifts due to the orbital motion of the plasma and by gravitational redshifts that increase at smaller radii, closer to the black hole. This broadens the narrow emission line, forming a blueshifted peak and redshifted wing [16].

### X-ray echoes from behind the black hole

When reflection is observed from the inner parts of an accretion disk, illuminated by a variable corona, we expect to see reverberation time delays, where variations in the reflection lag those in the primary X-ray emission due to the light travel time between the X-ray source and the disk. Such time lags have been measured in I Zw 1 [9]. We can combine the 2020 *XMM-*

Newton observations with the archival observations to obtain a measurement of the average reverberation timescale. We measure an iron K reverberation lag of $(746 \pm 157)$ s. While the structure of the corona is complex, the rapid variability is dominated by the compact core and thus we adopt a simplified model in which the rapidly variable X-ray emission originates from a compact, point-like source [12]. Given the best-fitting parameters of the X-ray spectrum, the time lag corresponds to a corona height of $4.3^{+1.7}_{-1.1} r_g$ above the accretion disk, where the gravitational radius, $r_g = GM/c^2$ is the radial position of the event horizon in the equatorial plane of a maximally spinning black hole.

Emission line photons from different parts of the disk experience different Doppler shifts, due to the variation in the line-of-sight velocity across the disk, and also experience gravitational redshifts, which increase closer to the black hole. The energy shifts of the line photons therefore contain information about the positions on the accretion disk from which they were emitted. The light travel time varies according to the distance of each part of the disk from the corona, and the line emission at different energy shifts is expected to respond to the flare at different times [11].

On the decline of each flare, we find a series of short peaks in the flux of, first, the blueshifted iron K line photons, then, at a later time the redshifted line photons (Figure 1b). These offset peaks are significantly detected, at 99.99 per cent confidence, with less than 0.01 per cent probability that they arise due to Poisson noise or random red noise variations in the observed light curves (see Methods). Similar peaks are detected during both flares, and we find that the series of three offset peaks occur at the same time after the onset of each flare, and the flares brighten through a similar mechanism (though the second flare decays more slowly).

We find that the offset timing of these peaks can be explained by a simple model of the reverberation of the iron K fluorescence line from the accretion disk during the flares (Figure 2). Figure 3 shows the response of the iron K emission line as a function of time and energy following a single flare of continuum emission, along with the response as a function of time in each of the observed energy bands. The earliest response is seen from the inner disk. Blueshifted emission arises from the approaching side of the disk while redshifted emission arises from the receding side. The most redshifted photons, reflected from the innermost radii on the disk, are delayed as they travel through the strong gravitational field. At late times, the photons from the outer disk are seen, close to 6.4 keV, where the Doppler shifts and gravitational redshifts are small. We find that the observed peaks correspond to the late-time edge of the 'loop' in the response function running from the blueshifted to the redshifted side of the line, shown by the blue lines in Figure 1(b). This track represents line emission that reverberates from the back side of the accretion disk that would be hidden behind the black hole. These rays are delayed and bent into the line of sight in the strong gravitational field close to the black hole, and are magnified by gravitational lensing. The peaks correspond to the times at which the regions of the back side of the disk that contribute to the response intersect the caustics projected by the black hole gravitational lens [11, 12].

## Extracting the reverberation response of the accretion disk

To investigate further this interpretation, we investigate the short-timescale variability in the shape of the X-ray spectrum. We compute the residuals of the spectrum after the best-fitting continuum has been subtracted (Figure 4a) and detect significant variability in both the soft X-rays (reflected from the disk) and the redshifted part of the iron K line from the inner disk. We see little variation in the residuals between 1 and 2 keV (most strongly dominated by the continuum) until around 7000 s, where we see the redshifted iron K emission reach these low energies, following a 'loop' feature between 3 and 7 keV. This feature corresponds to the 'loop' and 'waterfall' in the theoretical response of the iron line shown in Figure 3, where the re-emerging reflection from the back side of the disk makes up the late-time half of the loop and the low energy waterfall. Since the waterfall feature appears at late times after the initial flare, it is only visible in response to a particularly bright flare from the corona, otherwise it becomes smeared out and lost into the underlying continuum variability. These are the brightest short-duration flares (relative to the baseline luminosity) observed from I Zw 1, with the X-ray count rate increasing by a factor of 2.5.

The residual spectrum in successive time bins is shown in Figure 4(b). We find that the centroid of the iron K line begins redshifted with respect to the 6.4 keV rest-frame energy of the line (since the emission from the inner disk, closest to the corona, is redshifted), shifting up to 6 keV at 5600 s, in agreement with the prediction of the reverberation model. The re-emergence from the back side of the disk is then seen as the line brightens (due to gravitational lensing) and shifts to become increasingly redshifted at late time. The observed shift in the centroid of the iron K emission line during the flare, while predicted by models of X-ray reverberation from the accretion disk, cannot readily be explained by other models of the variability, discussed in the Supplementary Material.

## Properties of the corona and black hole

Fitting the reverberation model to the energy-resolved light curves, averaged across the two flares, we are able to make a second measurement of the height of the primary X-ray source above the disk, $h = 3.7^{+1.1}_{-0.7}\,r_\mathrm{g}$ and the mass of the black hole, $M_\mathrm{BH} = 3.1^{+0.5}_{-0.5} \times 10^7 M_\odot$. The model fits the reverberation response to the peaks and incorporates no prior assumptions about the height of the X-ray source. Yet, we find the height of the corona to be in agreement with the average coronal height inferred from the reverberation measurement. We conclude that the observed peaks are naturally expected from a short X-ray flare emitted from the corona in I Zw 1.

We note, however, that the appearance of such narrow X-ray re-emergence peaks requires the duration of the flare that illuminates the disk to be shorter than the timescale of the total response from the disk. The total duration of the flare is 10ks. The model requires that the disk only respond to the first part of the flare, prior to 4500 s, for the narrow peaks to appear at 6800 s without being smeared out. This requirement is not unreasonable. We observe a drop in the reflection fraction, which can be interpreted in terms of the corona accelerating away from the disk during the flares [17], thus a strong reverberation response is only seen to the initial part of the flare when a significant fraction of the continuum emission illuminates the inner disk.

Short-timescale X-ray flares from the corona in I Zw 1 reveal, for the first time, the temporal response of the illuminated accretion disk. We find the first direct observational evidence for the re-emergence of emission from behind the black hole, bent into our line of sight by strong gravitational light bending. These new observations confirm a key prediction of General relativity; the detection of photons bent around the black hole from the back side of the disk.

## References

[1] A. A. Galeev, R. Rosner and G. S. Vaiana, "Structured coronae of accretion disks," *ApJ,* vol. 229, pp. 318-326, April 1979.

[2] F. Haardt and L. Maraschi, "A two-phase model for the X-ray emission from Seyfert galaxies," *ApJ,* vol. 380, p. L51, October 1991.

[3] A. Merloni and A. C. Fabian, "Thunderclouds and accretion discs: a model for the spectral and temporal variability of Seyfert 1 galaxies," *MNRAS,* vol. 328, pp. 958-968, December 2001.

[4] I. M. George and A. C. Fabian, "X-ray reflection from cold matter in active galactic nuclei and X-ray binaries," *MNRAS,* vol. 249, pp. 352-367, March 1991.

[5] P. Uttley, E. M. Cackett, A. C. Fabian, E. Kara and D. R. Wilkins, "X-ray reverberation around accreting black holes," *A&ARv,* vol. 22, p. 72, 2014.

[6] A. C. Fabian et al., "Broad line emission from iron K- and L-shell transitions in the active galaxy 1H0707-495," *Nat.,* vol. 459, pp. 540-542, May 2009.

[7] T. Boller, W. N. Brandt and H. Fink, "Soft X-ray properties of narrow-line Seyfert 1 galaxies," *A&A,* vol. 305, p. 53, 1996.

[8] L. C. Gallo, "Investigating the nature of narrow-line Seyfert 1 galaxies with high-energy spectral complexity," *MNRAS,* vol. 368, pp. 479-486, May 2006.

[9] D. R. Wilkins et al., "Revealing structure and evolution within the corona of the Seyfert galaxy I Zw 1," *MNRAS,* vol. 471, pp. 4436-4451, November 2017.

[10] L. C. Gallo, W. N. Brandt, E. Costantini and A. C. Fabian, "A longer XMM-Newton look at I Zwicky - 1. Distinct modes of X-ray spectral variability," *MNRAS,* vol. 377, pp. 1375-1382, May 2007.

[11] C. S. Reynolds, A. J. Young, M. C. Begelman and A. C. Fabian, "X-Ray Iron Line Reverberation from Black Hole Accretion Disks," *ApJ,* vol. 514, pp. 164-179, March 1999.

[12] D. R. Wilkins, E. M. Cackett, A. C. Fabian and C. S. Reynolds, "Towards modelling X-ray reverberation in AGN: piecing together the extended corona," *MNRAS,* vol. 458, pp. 200-225, May 2016.

[13] F. A. Harrison et al., "The Nuclear Spectroscopic Telescope Array (NuSTAR) High-energy X-Ray Mission," *ApJ,* vol. 770, p. 103, June 2013.

[14] F. Jansen et al., "XMM-Newton observatory. I. The spacecraft and operations," *A&A,* vol. 365, p. L1, January 2001.

[15] G. Risaliti et al., "A rapidly spinning supermassive black hole at the centre of NGC 1365," *Nat.,* vol. 494, no. 7438, pp. 449-451, February 2013.

[16] A. C. Fabian, M. J. Rees, L. Stella and N. E. White, "X-ray fluorescence from the inner disc in Cygnus X-1," *MNRAS,* vol. 238, pp. 729-736, May 1989.

[17] D. R. Wilkins et al., "Flaring from the supermassive black hole in Mrk 335 studied with Swift and NuSTAR," *MNRAS,* vol. 454, pp. 4440-4451, December 2015.

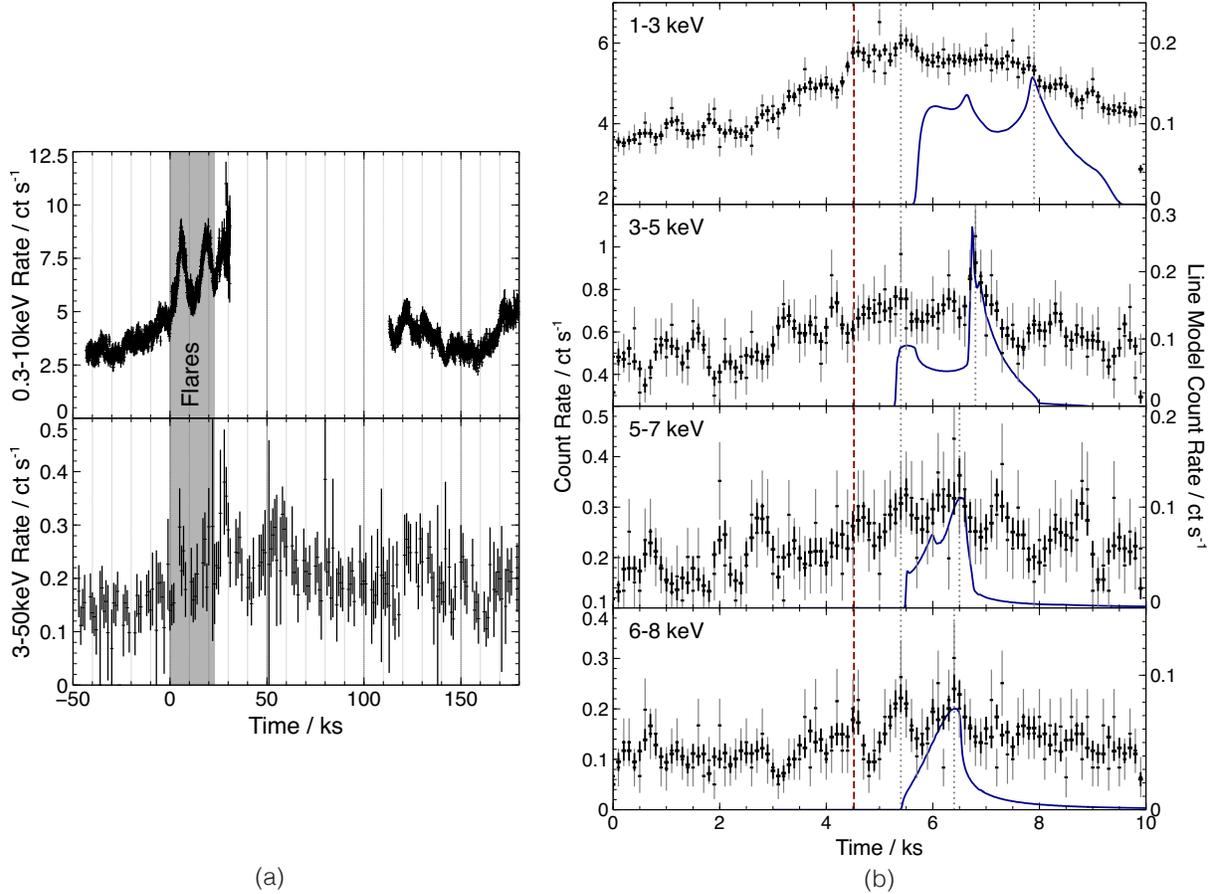

Figure 1: X-ray light curves of the AGN I Zw 1. (a) The total photon count rate measured by *XMM-Newton* in the 0.3-10 keV energy band, in 100 s time bins, (top panel), and by *NuSTAR* in the 3-50 keV energy band, in 1000 s time bins (bottom panel). X-ray flaring was observed toward the end of the first segment of the *XMM-Newton* observation. (b) X-ray light curves, summed over the two flares, in energy bands corresponding to the continuum-dominated 1-3 keV band, the redshifted wing of the iron K fluorescence line, emitted from the inner regions of the accretion disk (3-5 keV), the core of the line (5-7 keV), and the blueshifted iron K line photons from the approaching side of the disk (6-8 keV). Error bars represent 1σ uncertainties. As well as peaks that occur simultaneously between the energy bands, between 6.4 and 6.8 ks, short peaks or flashes are seen offset in time between the energy bands. These peaks are consistent with the re-emergence of photons from the flare, reflected from the part of the accretion disk behind the black hole, bent into the line of sight by the strong gravitational field. These peaks are described by the model of the excess iron K line emission (described in Methods), shown by the blue curve, responding to the initial part of the continuum flare, indicated by the red dashed line.

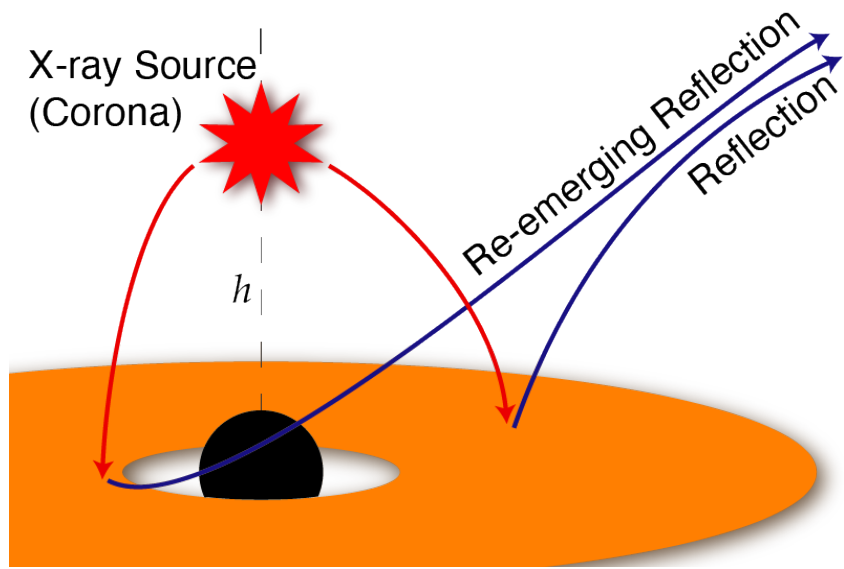

Figure 2: Schematic of the X-ray reverberation model. X-rays are emitted from a corona of energetic particles close to the black hole. While some of these reach the observer directly, a portion of these rays illuminate the inner regions of the accretion disk and are observed reflected from the disk. Strong light bending in the gravitational field around the black hole focuses the rays towards the black hole and onto the inner regions of the disk. Rays reflected from the back side of the disk can be bent around the (spinning) black hole allowing the 're-emergence' of X-rays from parts of the disk that would classically be hidden behind the black hole.

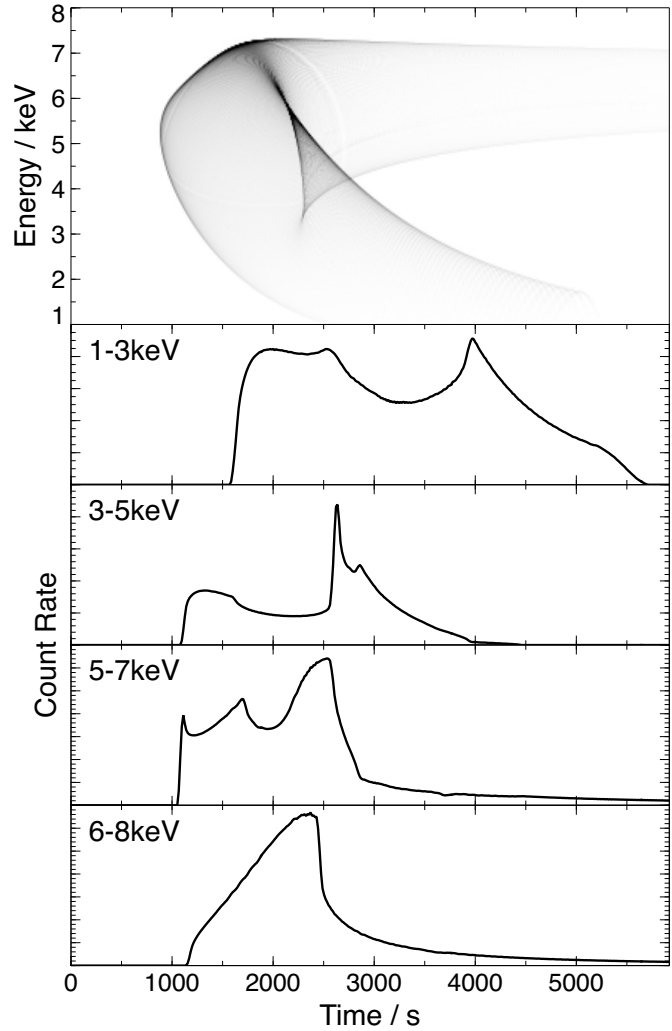

Figure 3: **Modelled response of the reverberating iron K fluorescence line to a flare of continuum emission.** The panels show the response of the reverberating iron K fluorescence line as a function of time following a single short flare of continuum emission from a point-like corona $4r_g$ above the disk. Time lags scale linearly with mass and are shown for a black hole with mass $3 \times 10^7\ M_\odot$. The top panel shows the count rate of line photons as a function of time and energy following the flare. The line is emitted at 6.4 keV in the rest frame of the emitting material, though the observed photon energies are shifted by Doppler shifts and gravitational redshifts from the orbiting material around the black hole. The lower panels shown the count rate in the specific energy bands measured in I Zw 1 as a function of time, showing the form of the reverberating emission line in different energy bands.

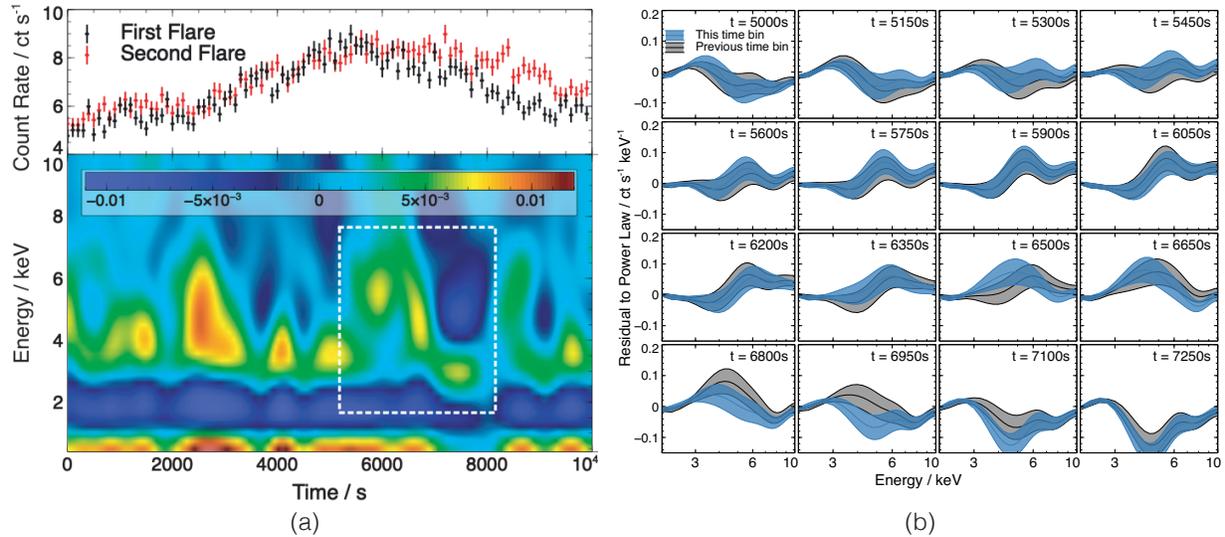

Figure 4: Features of the reverberation response function detected during the X-ray flares. (a) Map of the residuals, as a function of energy and time, of the observed spectrum in 50 s time bins, to the best fitting power law continuum in each time bin. The map is summed across the two flares and smoothed by a 2-dimensional Gaussian filter to suppress the effects of Poisson noise. Shading represents the difference between the observed and continuum model count rates (in ct s$^{-1}$ keV$^{-1}$). The residuals show variable reflection from the accretion disk. The characteristic response of the X-ray flare reverberating from the accretion disk is seen between 5000 and 8000 s, indicated by the white box. The top panel shows, for reference, the 0.3-10 keV light curve, showing the evolution of the two flares. (b) The evolution of the iron K line in the (smoothed) residual spectrum in 150 s time bins. The spectra show how the line emission during the flare starts redshifted, as the inner disk closest to the corona responds, then the centroid then shifts to the 6.4 keV rest frame energy of the line as the outer disk responds. At late times, the line becomes more and more redshifted as the re-emerging reflection from the back side of the disk is seen, lensed around the black hole. In each time bin, the spectrum (shown in blue) is compared to that in the previous time bin (shown in grey). The shaded regions show the 1σ confidence interval.

# Methods

## Observations and data reduction

I Zw 1 was observed continuously by *NuSTAR* over 456 ks (5.3 days), between 2020 January 11 and 2020 January 16 (OBSID 60501030002). The total exposure obtained was 233 ks, accounting for gaps in the observation every 90 minutes as the spacecraft, in low-Earth orbit, passed behind the Earth. Simultaneous observations were obtained by *XMM-Newton* during two continuous periods of 76 and 69 ks, beginning 100 and 260 ks into the *NuSTAR* observation (OBSIDs 0851990101 and 0851990201).

*NuSTAR* observations were reduced following the standard procedure, using NUSTARDAS v1.9.2. The event lists from each of the focal plane module (FPM) detectors were screened and reprocessed using the NUPIPELINE task, applying the most recent calibration available at the time of writing. Source photons were extracted from a circular region, 30 arcsec in diameter, centered on the point source. The smaller 30 arcsec extraction region, suitable for fainter hard X-ray sources, was selected over the larger 60 arcsec region to maximize the ratio of source to background count rate, owing to the steeply falling hard X-ray spectrum of I Zw 1 limiting the photon count above 10 keV. The background was extracted from a region, the same size, away from the point source on each detector. Source and background spectra, along with the corresponding response matrix and ancillary response (effective area) were extracted using the NUPRODUCTS task, in addition to source and background light curves both over the full bandpass and in specific energy bands. NUPRODUCTS automatically applies all appropriate dead time and exposure corrections to the extracted light curves.

The separate *NuSTAR* spectra obtained from the FPMA and FPMB detectors were examined for consistency. The best-fitting photon index describing the slope of the spectra was found to be consistent within the statistical uncertainty between the spectra from the two detectors, thus the FPMA and FPMB spectra were summed in order to maximize the signal-to-noise in the highest energy X-ray bands, using the average response matrices from the two detectors. I Zw 1 was significantly detected above the background by *NuSTAR* up to 50 keV.

*XMM-Newton* observations were reduced using the XMM SCIENCE ANALYSIS SYSTEM (SAS) v18.0.0. We here focus on the data collected by the EPIC pn camera [18], owing to its superior sensitivity, particularly for X-ray timing and variability studies. Event lists were reprocessed and filtered using the EPPROC task, using the latest version of the calibration. Source photons were extracted from a 35 arcsec region centered on the point source, while background photons were extracted from a region the same size, located on the same chip. Spectra were extracted using the EVSELECT task, while the response matrix and ancillary response were generated using the RMFGEN and ARFGEN tasks. After extraction by EVSELECT, light curves were corrected to account for dead time and exposure variations using the EPICLCCORR task.

## Modelling the X-ray spectrum

We began by examining the time-averaged 3-50 keV X-ray spectrum of I Zw 1 obtained by *NuSTAR* and *XMM-Newton* over the full duration of the observation, shown in Extended Data Figure 1a. The ratio of the observed X-ray spectrum to the best-fitting power law (Extended Data Figure 1b). The spectrum reveals the X-ray emission that is reflected by the inner accretion disk; the relativistically broadened iron K fluorescence line, centered at 6.4 keV with an extended, redshifted wing extending to 3 keV, in addition to the Compton Hump around 25 keV.

We therefore modelled the time-averaged spectrum by the combination of the directly observed continuum emission, described by a power law that is exponentially cut-off above an energy, $E_{cut}$, that represents the temperature of the emitting corona, and the relativistic reflection spectrum that is produced when plasma orbiting in the accretion disk is irradiated by the X-ray continuum. The reflection spectrum is modelled using RELXILL [19], which convolves the spectrum emitted by the irradiated plasma in its own rest frame from XILLVER [20, 21, 22] with the RELCONV kernel [23] to apply relativistic broadening due to gravitational redshifts and Doppler shifts from the orbiting accretion disk.

We applied the model firstly to the full *NuSTAR* observation in the 3-50keV energy range, and then simultaneously to the to the *XMM-Newton* EPIC pn and summed *NuSTAR* FPM spectra over the 3-50keV energy range (using only the parts of the *NuSTAR* observation simultaneous with the *XMM-Newton* observations). Including *XMM-Newton* data alongside *NuSTAR* maximizes the spectral resolution and photon count in the 3-50 keV band. We allow for an offset in the normalization of the spectra obtained by the two satellites due to calibration uncertainties via a cross-normalization constant that is fit to the observed spectra.

At X-ray energies below 1 keV intrinsic absorption has previously been found to significantly impact the X-ray spectrum of I Zw 1. From the measured column density and ionization state of the absorbing gas [24, 25], these absorbers are not expected to, and are not seen to, imprint significant features on the spectrum above 3 keV that can be detected with the limited spectral resolution of *NuSTAR* and the *XMM-Newton* EPIC pn camera. Below 1 keV, the X-ray spectrum becomes dominated by a soft excess of X-ray emission reprocessed by the accretion disk, which is sensitive to not just the ionization but the density [26] and structure [27] of the accretion disk that are not fully described by the RELXILL model. We therefore fit the spectral model to the 3-50keV band in order to obtain the cleanest measurements of the reflected X-rays from the inner disk.

Model parameters that best describe the observed spectrum were fit by minimizing the modified version of the Cash statistic (calibrated to approximate $\chi^2$ when there are least five counts per bin) using XSPEC. After the best fitting values of the model parameters were found, uncertainties were estimated from the posterior probabilities of each parameter, obtained using a Markov Chain Monte Carlo (MCMC) calculation. The model has 10 free parameters. Markov chains were produced using the Goodman-Weare algorithm with 80 walkers, for 50000 iterations after discarding the first 5000 iterations to remove 'memory' of the starting values.

The RELXILL model provides a good fit to the 3-50keV *XMM-Newton* and *NuSTAR* spectra, yielding a total C-statistic of 1067 (for 1098 degrees of freedom). There are no significant residuals, except for narrow absorption and emission features around 9keV that may represent an ultrafast outflow; a high ionization wind outflowing from the inner disk at 0.25c. Including Gaussian models for these features decreases the C-statistic by only 6 and does not affect the best-fitting parameters of the inner disk reflection spectrum.

The best-fitting parameters and their uncertainties are shown in Extended Data Table 1. By simultaneously fitting the model to the *XMM-Newton* and *NuSTAR* spectra over the 3-50 keV energy range, we are able to constrain the slope of the continuum spectrum emitted from the corona, the iron abundance and ionization parameter of the accretion disk and the inclination of the disk to the line of sight. We are able to constrain the spin of the black hole, with the extremal redshift detected in the redshifted wing of the iron K line requiring a black hole spin parameter $a = J/Mac > 0.75\ GM/c^2$. We note that from the X-ray spectrum the reflection fraction, defined as the ratio of the total observed reflected to continuum flux, is less than unity showing that the inner regions of the accretion disk are under-illuminated compared to the expectation of illumination by a static point source. Dividing the observations into time segments before, during and after the flares and fitting the model to each segment (tying the values of the iron abundance, inclination and black hole spin parameter, which should not vary, between the time segments), we find that the reflection fraction drops from $0.68^{+0.22}_{-0.17}$ before the flares, to $0.16^{+0.11}_{-0.09}$ during the flares, rising again to $0.45^{+0.37}_{-0.04}$ afterwards.

We model the emissivity profile of the accretion disk (the reflected flux as a function of radius) using a twice-broken power law, motivated by general relativistic ray tracing models of accretion disks illuminated by either a compact point source or a corona that extends over the inner disk [28]. The best-fitting reflection model shows an emissivity profile falling steeply, following $r^{-8}$ over the innermost parts of the disk, consistent with the X-ray emission from the corona being focused towards the black hole in the strong gravitational field. The emissivity profile then flattens out to a radius of $15r_\mathrm{g}$, before falling off as approximately $r^{-3}$ over the outer disk, in agreement with the predictions of a model in which the corona consists of a bright, rapidly variable core in addition to a component that extends over the inner parts of the disk [9].

## X-ray reverberation

The *XMM-Newton* observations obtained in 2020 are too short to obtain a high signal-to-noise measurement of the time lags between different energy bands responding to a change in flux. They can, however, be combined with the archival observations of I Zw 1 (OBSIDs 0110890301, 0300470101, 0743050301 and 0743050801) to measure X-ray reverberation time lags between variability in the continuum emission and the reflection from the disk, and to determine the average scale height of the corona above the disk.

We extracted light curves in 10 approximately logarithmically spaced energy bands between 0.3 and 10 keV, with 10 s time binning and compute, from their Fourier transforms, the cross-spectrum of each with respect to a reference band [5]. The reference band is the sum of all of the energy bands, except for the energy band in question, so as to maximize signal-to-noise in the reference band, while avoiding correlated noise between the bands. The cross spectrum is a function of Fourier frequency, describing the slow and fast components that make up the observed variability. The time lag (derived from the phase of the cross spectrum) represents the average response time of each energy band to variations on each timescale, relative to the reference band. We find that reverberation time lags, where the soft excess and broad iron K line from the accretion disk lag behind the continuum-dominated 1-2 keV band, are detected with the highest signal-to-noise over the Fourier frequency range 5 to $7 \times 10^{-4}$ Hz, in agreement with the frequency range over which the reverberation was previously detected in I Zw 1 [9]. The time lag as a function of energy is shown in Extended Data Figure 2. The magnitude of the iron K lag (from the continuum band to the peak of the line) is measured to be $(746 \pm 157)$ s.

In order to estimate the height of the corona from the time lag measurement, it is necessary to account not only for the light travel times in the curved space time around the black hole (which are delayed relative to straight light paths in flat space), but also the spectral dilution of the observed lags. The bands dominated by the reflection from the disk will contain a contribution from the directly observed continuum emission and vice versa, reducing the measured lags by a factor of a few [29]. We construct a model of X-ray reverberation from a simplified, point-like corona, convolving the reverberation response function (see section on Reverberation Modelling below) with the best-fitting reflection spectrum model. With this reflection model, the measured time lag suggests reverberation from a corona at height $4.3^{+1.7}_{-1.1} r_g$.

## Detection and significance of peaks in the flare light curves

In order to investigate the reverberation of the X-ray flares off the accretion disk, light curves were extracted from the *XMM-Newton* EPIC pn observations in energy bands corresponding to the iron K emission, shifted by gravitational redshifts and Doppler shifts from different parts of the disk. The EPIC pn data were selected due to the instrument's superior collecting area, sensitivity and timing resolution and the light curves were analyzed individually to avoid systematic errors that may arise from combining data from different instruments. Light curves were extracted in 100s time bins in the following energy bands, with the width selected to maximize the signal to noise in each band, minimizing the contribution of Poisson noise to the variability:

- 1-3 keV, dominated by directly observed continuum emission, along with only the most highly redshifted iron K line photons from the innermost disk.
- 3-5 keV, dominated by redshifted iron K line photons (the wing of the emission line), from the inner regions of the disk.
- 5-7 keV, corresponding to the core of the iron K line, encompassing moderate redshifts as well as the line photons seen from the outer disk where the Doppler shifts and gravitational redshifts are small.
- 6-8 keV, showing the blueshifted iron K line photons, from the approaching side of the inner disk.

The high frequency variability is dominated by Poisson noise, characterized by a constant power spectral density in frequency. In the *XMM-Newton* light curves, we find the power spectrum to be constant and hence dominated by Poisson noise above $10^{-3}$ Hz. In order to suppress the high frequency variability that is due to Poisson noise in the plotted light curves, we therefore apply a three-point moving average filter (a low-pass filter), suppressing the high frequency components of the variability in the observed light curve, while maintaining intrinsic variability on longer timescales. In order to compute the uncertainty in the filtered light curves, each light curve was resampled 10,000 times by drawing the count rate in each time bin from a Poisson distribution with mean equal to the observed count rate. The distribution of the count rate in each time bin after applying the moving average filter was obtained and the uncertainty of each time bin was the standard deviation of the resampled count rates.

**Significance of peaks with respect to Poisson noise**

To assess the significance with which the narrow peaks are detected in the iron K band light curves, we aim to test the null hypothesis that the peaks in the raw light curves (i.e., with no moving average filter applied) are due to random variations in the count rate due to Poisson noise via Monte Carlo simulation. We estimate the baseline count rate above which the peaks are detected by applying a 25-point moving average filter to the observed light curves to produce a smoothed time series with no such short-timescale variations. From the smoothed light curve in each energy band, we generate 1,000,000 sample light curves, drawing each time bin from a Poisson distribution, and we compare the observed light curves to this sample.

Firstly, we assess the probability of each time bin in the peaks reaching the measured count rate in excess of the baseline variations. We compute the number of sample light curves in which the peak time bins reach or exceed the observed level and find that we can reject the null hypothesis of these time bins appearing as peaks in the 3-5keV band at the 99.99 per cent confidence level, in the 5-7keV band at 99.4 per cent confidence and in the 6-8keV energy band at 99.8 per cent confidence. The significance of the peaks is visualized in Extended Data Figure 3(a), which shows the improvement in the $\chi^2$ statistic over describing the observed light curves using the 25-point smoothed baseline light curve, when a Gaussian peak is added in each time bin, with varying normalization. In each case, the Gaussian peak is allowed to have variable width that is fit to the data (note, however, that we use Monte Carlo simulations, rather than this $\chi^2$ statistic to test the significance of each peak).

We must consider the probability of not just a single time bin appearing above the baseline count rate, but the probability of any of the time bins reaching this level, i.e. there have been multiple trials [30]. In each energy band, the light curve exceeds the baseline in three consecutive 100s bins during the peaks. We therefore assess the frequency at which three consecutive time bins exceed the baseline light curve by the observed factors at any point within the light curve. We reject the null hypothesis that the peaks appear by random Poisson fluctuations from the baseline light curve in the in the 3-5keV energy band at the 99.998 per cent confidence level, in the 5-7keV band at 99.2 per cent confidence and in the 6-8keV band at 99.5 per cent confidence. We therefore conclude that the narrow peaks, offset in time between the redshifted 3-5keV and blueshifted 6-8keV are significantly detected and the broader peak in the 5-7keV band is less significantly detected.

The bottom panel of Extended Data Figure 3(a) illustrates the time bins in which peaks are most significantly detected in each light curve. We see the same pattern of offset peaks on the decline of the second flare as on the decline of the first. The peak is seen first in the 6-8 keV light curve, and then in the 3-5 keV 500 s later. We note, however, that the peaks are weaker in the second light curve, being detected at the 99.96, 99.8 and 98 per cent confidence levels in the 3-5 keV, 5-7 keV and 6-8 keV bands.

Peaks are also detected at the maximum of the first flare, 5.4 ks from the start of the light curve (Figure 1b). These peaks are detected simultaneously across all energy bands and represent rapid variability in the broadband X-ray continuum emission that peaks sharply and is not represented by the smoothed baseline model. We also see a peak in the soft X-ray emission as the second flare begins, at 15.3 ks. This peak is seen simultaneously in the 3-5 keV and 5-7 keV bands, but is not significantly detected in the 6-8 keV band. The simultaneity between the 3-5 and 5-7 keV bands suggests that this is also variability in a broadband continuum component, but its non-detection at 6-8 keV suggests that this emission component peaks at low energy.

We sum the light curves from the two flares, aligning the two light curves by the timing of the 3-5 keV peak. Remarkably, we find that when the time series are aligned, the rising profile of the two flares is almost identical. The flares appear to begin by the same mechanism, and the series of offset peaks across the energy bands are common to the two flares and occur the same time after the beginning of the flares. Stacking the light curves in this way suppresses many of the random fluctuations and increases the confidence level to which the flares are detected to 99.9999 per cent in the 3-5 keV band, 99.94 per cent in the 5-7 keV band, and 99.97 per cent in the 6-8 keV band (Extended Data Figure 3b).

**Significance of peaks with respect to random red noise variability**

It is also possible that the peaks appearing at different times in the different light curves could have arisen due to random red noise variability that is uncorrelated between the energy bands. A high degree of coherence, however, is measured, placing an upper limit on the level of uncorrelated variability that can exist between the light curves in any two energy bands. The coherence measures the fraction of the variability in one energy band that can be predicted by linear transformation of that in a second energy band [5]. Between the 3-5 keV energy band, dominated by the redshifted wing of the iron line, and the 1-3 keV energy band, dominated by the continuum, we measure the coherence to be as high as 0.94 up to a frequency of $2 \times 10^{-4}$ Hz (Extended Data Figure 4).

In order to determine the fraction of variability that can be uncorrelated, given this level of coherence, we simulate a further 100,000 pairs of random light curves which possess a power spectrum falling with frequency as $f^{-2}$, but reproducing the probability distribution of the count rates observed in the light curves we observe from I Zw 1 [31]. For each pair of light curves, we generate three random time series, one which describes the correlated variability between the two bands, $L_{\text{corr}}$, and two that represent the uncorrelated variability in each band, $L_A$ and $L_B$. Each of these time series are normalized to have a mean count rate of zero (allowing negative values at this stage) and standard deviation of unity. We then form the simulated pair of 'observed' light curves from combinations of these such that a fraction u of the variability is uncorrelated:

$$L_1 = \sqrt{1 - u^2} L_{\text{corr}} + u L_A,$$

$$L_2 = \sqrt{1 - u^2} L_{\text{corr}} + u L_B$$

The resulting light curves are then rescaled to possess the same mean and standard deviation, and approximately the same probability distribution of count rates, as the observed light curves. Poisson noise is then added to each of the simulated light curves.

We find that in order to maintain coherence as high as 0.94 at $2 \times 10^{-4}$ Hz, it is necessary that $u < 0.1$ (Extended Data Figure 4). This means that almost all of the observed variability is correlated between the observed energy bands and can be described by a linear transformation of the continuum variability through the reverberation response function. The observed drop in coherence at high frequency can almost entirely be attributed to Poisson noise.

Using these same pairs of light curves, we assess the probability that the peaks observed at different time bins in different energy bands, which we interpret as the re-emergence of reflected X-rays from the far side of the accretion disk, could appear by chance from random, uncorrelated red noise variations. We simulate one sample of 10,000 light curve pairs that represent the 3-5 keV band relative to the 1-3 keV band, and a further sample of 10,000 light curve pairs that represent the 6-8 keV band relative to the 1-3 keV band. Using the same peak-detection criterion described above, we compute the probability that the feature we detect in the observations of I Zw 1 arises due to random variations, i.e. that a short peak appears in the 3-5 keV band (above the underlying slower variability in the smoothed light curve), which is not accompanied by a peak in the 1-3 keV band (which has the highest count rate and, thus, the lowest fraction of its variability due to Poisson noise), while in a separate time bin, a peak appears in the 6-8 keV band, also unaccompanied by a peak in the 1-3 keV band.

We determine that the probability of such a series of peaks arising due to the combination of uncorrelated red noise variability and Poisson noise during just one of the observed flares is 0.009 per cent; thus, the feature is detected in the light curves at greater than 99.99 per cent confidence.

## Visualization of the reverberation response function

Having determined that the series of offset peaks upon the decline of the flares in different energy bands across the relativistically broadened iron K line are significantly detected, we further explore their origin. We use the measured light curves to make the first attempt to recover the reverberation response function from the data; that is the reflected flux as a function of energy and time from the onset of each flare.

We extract light curves from the first *XMM-Newton* observation in 40 linearly spaced energy bins between 0.3 at 10 keV, and in 50 s time bins. From these, we create a grid of count rate per energy per time bin. For the visualization of the response function, we then apply a 2D Gaussian filter (with 4x4 bin standard deviation defining the width). The Gaussian filter acts to average the high frequency fluctuations that occur due to Poisson noise across neighboring bins, but unlike a moving average filter, preserves sharp features in the grid. In each time bin, we then fit the spectrum with an absorbed power law (with hydrogen column density corresponding to the Galactic column density along the line of sight to I Zw 1, $n_\mathrm{H} = 4.6 \times 10^{20}\,\mathrm{cm}^{-2}$). The model is folded through the *XMM-Newton* EPIC pn response matrix. The final product is then the 2D grid of the residual to this model in each time and energy bin. This method is similar to that used to produce excess residual plots [32], except extracting the residuals from light curves extracted by the instrument pipeline, rather than spectra, to achieve finer time binning. The grid of residuals shows what remains after the underlying continuum is subtracted. This includes both the primary continuum from the corona and the continuum component of the accretion disk reflection spectrum.

The time-evolution of the iron K line spectrum is extracted from successive time bins of this 2D residual grid. The uncertainty in the spectrum, given Poisson noise in the observed photon counts, is computed by Monte Carlo resampling the raw light curves that were used to construct the grid. For each raw light curve, the count rate in each time bin was drawn at random from a Poisson distribution with mean corresponding to the observed count rate. We then obtain the range of residual values for each grid point, over the sample of input light curves, and the uncertainty on each spectrum is shown as the standard deviation of each bin.

### X-ray reverberation light curve modelling

The reverberation of X-rays from the accretion disk was modelled using general relativistic ray tracing simulations, implemented using the CUDAKERR code [28, 12]. In these simulations, rays are emitted isotropically from a point source at rest on the rotation axis above the black hole and traced by integrating the null geodesic equations in the Kerr spacetime until they reached the accretion disk, assumed to lie in the equatorial plane. The accretion disk is assumed to be optically thick and geometrically thin, appropriate for the accretion flows around black holes accreting up to around 30 per cent of the Eddington limit [33]. A bolometric correction factor can be used to estimate the total, bolometric luminosity of I Zw 1 from the observed luminosity in the 2-10 keV X-ray band. From the measured luminosity of $4.5 \times 10^{43}\,\mathrm{erg\,s^{-1}}$, we can estimate that for a black hole mass of $2.8 \times 10^7\,\mathrm{M_\odot}$, I Zw 1 is accreting at 0.3 times the Eddington limit. Estimating the bolometric luminosity from the optical luminosity measured at 5100Å ($3.19 \times 10^{44}\,\mathrm{erg\,s^{-1}}$ [34]), however, we find I Zw 1 to be accreting at approximately the Eddington limit (noting that this could be an under-estimate due to 5100Å photons being trapped in the disk). The discrepancy arises from deviations between the specific spectral energy distribution of I Zw 1 and the model assumed in computing the bolometric correction factors. However, if I Zw 1 is accreting at this higher rate, radiation trapping in the disk will cause the disk to expand into a slim disk profile, which may also explain the detection of X-ray absorption in an outflowing wind. This will lead to slight deviations from the energy shifts in the emission lines that are predicted by this reverberation model.

For each ray, the time co-ordinate at which it reaches the disk is recorded along with its energy shift, $g$. The energy of photons along each ray is shifted by both the gravitational redshift and Doppler shift, assuming the material in the disk travels in a stable circular orbit at each radius. Rays are counted into radial bins on the disk and the intensity of the line emitted from each radius (in the co-rotating rest frame) is determined by the number of rays landing in each bin (which accounts for the aberration in solid angle between the source and the disk), weighted by $g^{1+\Gamma}$. This factor accounts for the shift in photon arrival rate due to the relative passage of proper time between the emitter and receiver, and the shift in the number of photons available in the received spectrum (with photon index $\Gamma$) above 7.1 keV required to excite the fluorescence line. Rays emitted from the disk are then traced to the observer (i.e. the telescope). This is achieved by setting up a regular grid of rays that travel perpendicular through a flat image plane some large distance away from the singularity that represents the observed patch of the sky. These rays are traced backwards until they reach the accretion disk, where again their travel time and energy shift $g$ are recorded. The intensity of the line photons, whose energies have been shifted, reaching the observer from each radius on the disk is further weighted by $g^3$ to account for the shift in photon arrival rate and solid angle aberration (using Liouville's theorem). The bending of rays as they propagate around the black hole, coupled with this weighting factor account for gravitational lensing and the corresponding magnification of regions of the disk behind the black hole that produce the narrow peaks in the reverberation response as photons that 're-emerge' from the far side of the disk.

X-ray reverberation is modelled via the impulse response function that is received as a function of energy and time after a single, short flash of continuum emission from the primary X-ray source. Ray tracing calculations are conducted in natural units in the gravitational field, measuring distances in gravitational radii, and times in units of $GM/c^3$, the light travel time over one gravitational radius. These units are straightforwardly converted into the observed times by multiplying by the mass of the black hole, a free parameter in the model.

The reverberation from the accretion disk was computed in response to a single flare of emission from the corona whose intensity as a function of time was described by a Gaussian function, for simplicity. The observed reverberation response is the Gaussian profile of the flare convolved with the impulse response function, summed over the energy bands that correspond to the observed light curves. The inclination of the accretion disk was taken to be 46 deg, as measured from the X-ray spectrum, and the accretion disk was assumed to extend inwards to the innermost stable of a maximally spinning black hole, at $1.235 r_g$.

We apply this model of X-ray reverberation from a single flash from a point-like corona simultaneously to the light curves, summed across the two flares, in the 3-5 keV and 6-8 keV bands, in which the narrow, offset peaks were most significantly detected. These bands correspond to, respectively, the redshifted wing and the blueshifted peak of the reverberating iron K fluorescence line. We do not include the 5-7 keV band in the analysis because the peak is less significantly detected in this light curve (it is broader so less easily distinguished from the background variability) and this energy band is not strictly independent of the 3-5 and 6-8 keV bands. This band, however, illustrates how the peak in the 5-7 keV band is observed at a time interim to that in the 3-5 and 6-8 keV bands, in agreement with the model. We fit the model count rate in each energy band and in each time bin during the period of the first flare by minimizing the Cash statistic, summing the statistic across the two light curves.

In order to apply this reverberation model to the observed X-ray flare, we begin by modelling the underlying variability in the light curve. To do this, we work from an initial assumption that the baseline light curve in each energy band is proportional to the total light curve measured in the *XMM-Newton* bandpass (that is to say that that the spectral shape remains approximately constant on short timescales during the flare, even though we know that the spectrum is variable, and that the spectrum during the flares differs from that before and after the flares — residuals to this model will represent short-timescale variability in the spectrum that we will model here). We initially model the light curve in each band as the baseline (broadband) light curve, multiplied by a free constant. To this baseline model, we add the model reverberation response. The model has five free parameters; the centroid time and width of the primary Gaussian continuum flare that is seen to reverberate, the mass of the black hole, the height of the point source and the normalization of the reverberation component. The reverberation parameters are tied between the energy bands and only the overall normalization of the reverberating component is fit. The relative normalization of the reverberation in each energy band is set by the line model. The model simultaneously describes the offset peaks at t = 6800s in the 3-5keV band and t = 6400 s in the 6-8keV band. The model does not account for the simultaneous narrow peaks that appear at all energies at the peak the flare (t = 4300 s), though the simultaneous nature of these peaks means that they can be readily explained by a short timescale variation in the continuum emission itself at the apex of the flare.

We step through values of the black hole mass and X-ray source height, using the Levenberg-Marquardt minimization algorithm finding the optimal values of the other parameters at each step. When considering just the average time lag between continuum photons and the reverberating line photons from the disk, the height of the X-ray source is degenerate with the black hole mass; increasing the black hole mass and increasing the height of the X-ray source (in units of $GM/c^2$) both increase the time lag. The detection of narrow peaks in the light curves, corresponding to the re-emergence of reflected photons from the behind the black hole, however, place further constraints on the height of the X-ray source. The amplitude of the peaks limits how high the X-ray source can be above the black hole; if it is too far from the disk, the innermost regions that are most strongly lensed are under-illuminated, and the amplitude of the peaks is decreased. Moreover, while the lag time from the continuum flare to the reverberation increases as the source height increases, the time between the re-emergence peaks in the redshifted and blueshifted light curves is only weakly dependent on the source height. These peaks are comprised of photons reflected from a narrow range of radii on the disk, thus the time lag between the peaks depends primarily on the distance around the disk from the blueshifted to the redshifted side, which depends on the black hole mass.

Modelling the X-ray reverberation from the start of the flare in the observed light curves, we are able to measure the height of the primary X-ray source above the disk, $h = 3.7^{+1.1}_{-0.7}\, r_g$ and the corresponding mass of the black hole $M_{\rm BH} = 3.1^{+0.4}_{-0.5} \times 10^7 M_\odot$ (where the errors correspond to the 90 per cent confidence interval). We find that centroid of the flare is at t = 3700s (corresponding to the initial rise in count rate seen during the flare) and that the disk responds just the first part of each flare to produce the narrow bright peaks.

### The mass of the black hole
The mass of the black hole measured from the X-ray reverberation model is consistent with that obtained from the width of the Hβ line in the optical spectrum, $M_{\rm BH} = 2.8^{+0.6}_{-0.7} \times 10^7 M_\odot$ [35], though we note that X-ray reverberation from the inner disk and measurement of the narrow re-emergence peaks in the light curve is more constraining of the lower mass bound, since such a low mass would require an X-ray source too high above the disk to produce strong peaks. Optical reverberation mapping using Hβ lines emitted from the broad line region finds the mass to be $M_{\rm BH} = 9.3^{+1.3}_{-1.4} \times 10^6 M_\odot$ [34], apparently in tension with both the mass derived from the line width and the best fitting value in our X-ray reverberation model. We note however that this analysis assumed the virial factor, which empirically represents the relationship between the black hole mass, line width and reverberation timescale, to be unity. This low value of the mass suggests accretion at a rate close to or slightly in excess of the Eddington limit, in which case the virial factor can be a factor of two to three higher [36]. It is therefore likely that optical reverberation mapping underestimates the mass of the black hole in I Zw 1 by a factor of two to three.

### Alternative explanations of the short-timescale variability
The observed shift in the centroid of the iron K emission line in the residual spectra (Figure 4) during the flare, while predicted by models of X-ray reverberation from the accretion disk, cannot readily be explained by other models of the variability. Variation in the X-ray continuum between time bins is accounted for when the residuals are computed, thus are not seen here. High column density, ionized outflows can reduce produce broad absorption features up to 2 keV, however variation in these outflows cannot explain the shift in the line centroid between 3 and 6 keV. Furthermore, outflows capable of producing such significant absorption would be inconsistent with the observed X-ray spectrum (the X-ray spectrum during the flare shows no evidence for significant absorption). Changes in the ionization of the accretion disk can change the rest frame energy of the iron K line between 6.4 and 6.97 keV and over-ionization of the inner disk can weaken the redshifted line emission seen between 3 and 5 keV relative to that at 6 keV but is likewise not able to produce the pattern of variability and the shift in the emission line centroid that is seen.

### References

[18] L. Strüder et al., "The European Photon Imaging Camera on XMM-Newton: The pn-CCD camera," *A&A,* vol. 365, p. L18, January 2001.

[19] T. Dauser et al., "Normalizing a relativistic model of X-ray reflection. Definition of the reflection fraction and its implementation in relxill," *A&A,* vol. 590, p. A76, May 2016.

[20] J. García and T. R. Kallman, "X-ray Reflected Spectra from Accretion Disk Models. I. Constant Density Atmospheres," *ApJ,* vol. 718, pp. 695-706, August 2010.

[21] J. García, T. R. Kallman and R. F. Mushotzky, "X-ray Reflected Spectra from Accretion Disk Models. II. Diagnostic Tools for X-ray Observations," *ApJ,* vol. 731, p. 131, April 2011.

[22] J. García et al., "X-Ray Reflected Spectra from Accretion Disk Models. III. A Complete Grid of Ionized Reflection Calculations," *ApJ,* vol. 768, p. 146, May 2013.

[23] T. Dauser, J. Wilms, C. S. Reynolds and L. W. Brenneman, "Broad emission lines for a negatively spinning black hole," *MNRAS,* p. 1460, September 2010.

[24] C. V. Silva et al., "The variability of the warm absorber in I Zwicky 1 as seen by XMM-Newton," *MNRAS,* vol. 480, no. 2, pp. 2334-2342, October 2018.

[25] E. Costantini et al., "A longer XMM-Newton look at I Zwicky 1: physical conditions and variability of the ionized absorbers," *MNRAS,* vol. 378, p. 873, 2007.

[26] J. A. García et al., "The effects of high density on the X-ray spectrum reflected from accretion discs around black holes," *MNRAS,* vol. 462, no. 1, pp. 751-780, 2016.

[27] C. Done, S. W. Davis, C. Jin, O. Blaes and M. Ward, "Intrinsic disc emission and the soft X-ray excess in active galactic nuclei," *MNRAS,* vol. 420, no. 3, pp. 1848-1860, 2012.

[28] D. R. Wilkins and A. C. Fabian, "Understanding X-ray reflection emissivity profiles in AGN: locating the X-ray source," *MNRAS,* vol. 424, pp. 1284-1296, August 2012.

[29] D. R. Wilkins and A. C. Fabian, "The origin of the lag spectra observed in AGN: Reverberation and the propagation of X-ray source fluctuations," *MNRAS,* vol. 430, no. 1, pp. 247-258, 2013.

[30] W. H. Press and P. Schechter, "Remark on the Statistical Significance of Flares in Poisson Count Data," *ApJ,* vol. 193, pp. 437-442, October 1974.

[31] D. Emmanoulopoulos, I. M. McHardy and I. E. Papadakis, "Generating artificial light curves: revisited and updated," *MNRAS,* vol. 433, no. 2, pp. 907-927, 2013.

[32] K. Iwasawa, G. Miniutti and A. C. Fabian, "Flux and energy modulation of redshifted iron emission in NGC 3516: implications for the black hole mass," *MNRAS,* vol. 355, no. 4, pp. 1073-1079, 2004.

[33] R. Narayan and E. Quataert, "Black Hole Accretion," *Science,* vol. 307, no. 5706, p. 77, 2005.

[34] Y.-K. Huang et al., "Reverberation Mapping of the Narrow-line Seyfert 1 Galaxy I Zwicky 1: Black Hole Mass," *ApJ,* vol. 876, no. 2, p. 102, 2019.

[35] M. Vestergaard and B. M. Peterson, "Determining Central Black Hole Masses in Distant Active Galaxies and Quasars. II. Improved Optical and UV Scaling Relationships," *ApJ,* vol. 641, pp. 689-709, April 2006.

[36] S. Collin, T. Kawaguchi, B. M. Peterson and M. Vestergaard, "Systematic effects in measurement of black hole masses by emission-line reverberation of active galactic nuclei: Eddington ratio and inclination," *A&A,* vol. 456, no. 1, pp. 75-90, 2006.

## Acknowledgements

This work was supported by the NASA *NuSTAR* and *XMM-Newton* Guest Observer programs under grants 80NSSC20K0041 and 80NSSC20K0838. DRW received additional support from a Kavli Fellowship at Stanford University. WNB acknowledges support from the V.M. Willaman Endowment. Computing for this project was performed on the Sherlock cluster. DRW thanks Stanford University and the Stanford Research Computing Center for providing computational resources and support. We thank the referees for their valuable feedback on the initial version of this letter.

## Author contributions

DRW performed the data analysis and reverberation modelling. LCG contributed to the analysis of the *XMM-Newton* spectra. WNB, EC and RDB contributed to the interpretation and discussion of the results.

## Data availability

The data used in this study from the *NuSTAR* and *XMM-Newton* are publicly available. *NuSTAR* observations can be accessed via the *NASA High Energy Astrophysics Science Archive Research Center* (https://heasarc.gsfc.nasa.gov). This work includes data obtained during *NuSTAR observation ID* 60501030002. *XMM-Newton* observations can be accessed via the *XMM-Newton Science Archive* (http://nxsa.esac.esa.int/nxsa-web). The primary analysis was conducted on observation IDs 0851990101 and 0851990201. Observation IDs 0110890301, 0300470101, 0743050301 and 0743050801 were included in the measurement of the average reverberation timescale.

# Extended Data

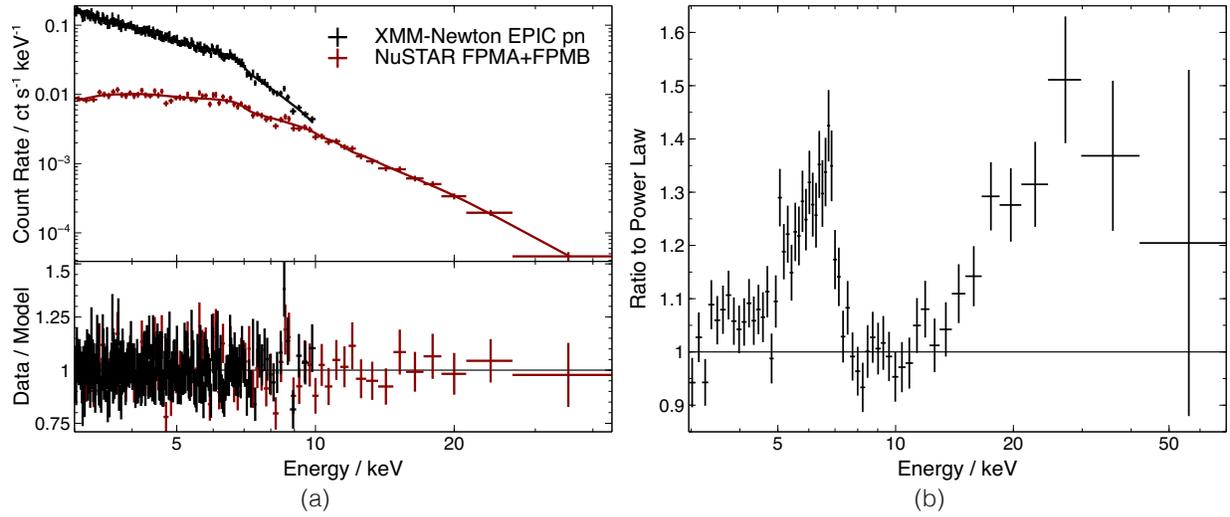

Extended Data Figure 1: The X-ray spectrum of I Zw 1. (a) The soft and hard X-ray spectra of I Zw 1 obtained by *XMM-Newton* and *NuSTAR* during the 2020 observations, fit with a model describing the emission in the 3-50 keV band as the combination of directly observed continuum emission and its reflection from the disk, described by RELXILL. (b) Ratio of the X-ray spectrum in the 3-78keV energy range measured with *NuSTAR* to the best-fitting power law continuum model, showing the residual features that arise due to the X-rays reflected from the inner accretion disk; the broad iron K line (around 6.4 keV) with redshifted wing, and the Compton hump centered at 25 keV. Error bars represent 1σ uncertainties due to Poisson noise.

Extended Data Table 1: Parameters of the X-ray continuum and reflection from the accretion disk. Parameters were measured by fitting the RELXILL model of relativistically blurred reflection from an accretion disk around a black hole that is illuminated by continuum emission from the corona. We fit the model simultaneously to the *XMM-Newton* and *NuSTAR* spectra in the 3-50 keV energy range. Uncertainties are shown in the 90 per cent confidence interval.

| | | |
|---|---|---|
| Continuum | Photon Index | $2.15^{+0.08}_{-0.06}$ |
| | Cut-off kT / keV | $> 102$ |
| Reflection | Reflection Frac. | $0.29^{+0.22}_{-0.10}$ |
| | Iron Abundance / Solar | $5^{+4}_{-3}$ |
| | Ionization ($\log \xi$ / erg cm s$^{-1}$) | $3.0^{+0.3}_{-2.6}$ |
| | Inclination / deg | $46^{+6}_{-5}$ |
| | Spin ($a$ / GM c$^{-2}$) | $> 0.75$ |
| Emissivity | Inner index | $8^{+1}_{-3}$ |
| | Inner break radius / $r_g$ | $4.3^{+0.2}_{-2.4}$ |
| | Middle index | $0.1^{+0.7}$ |
| | Outer break radius / $r_g$ | $15^{+27}_{-4}$ |
| | Outer index | $2.7^{+1.1}$ |

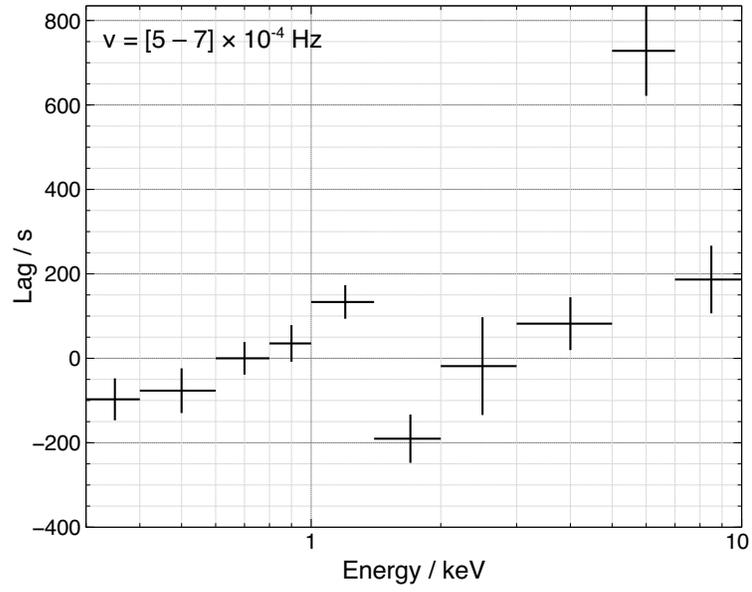

Extended Data Figure 2: **The relative response times of the X-ray emission at different energies to variations in luminosity.** The time lag is computed from the cross spectrum, averaged over Fourier frequency components from 5 to 7 x $10^{-4}$ Hz, from all *XMM-Newton* observations of I Zw 1 between 2002 and 2020. The lag vs. energy spectrum shows the delay of the response of the iron K line and soft X-ray excess, which reverberate from the accretion disk, with respect to the continuum which is most dominant in the 1-2 keV energy band. Error bars show the 1σ confidence interval.

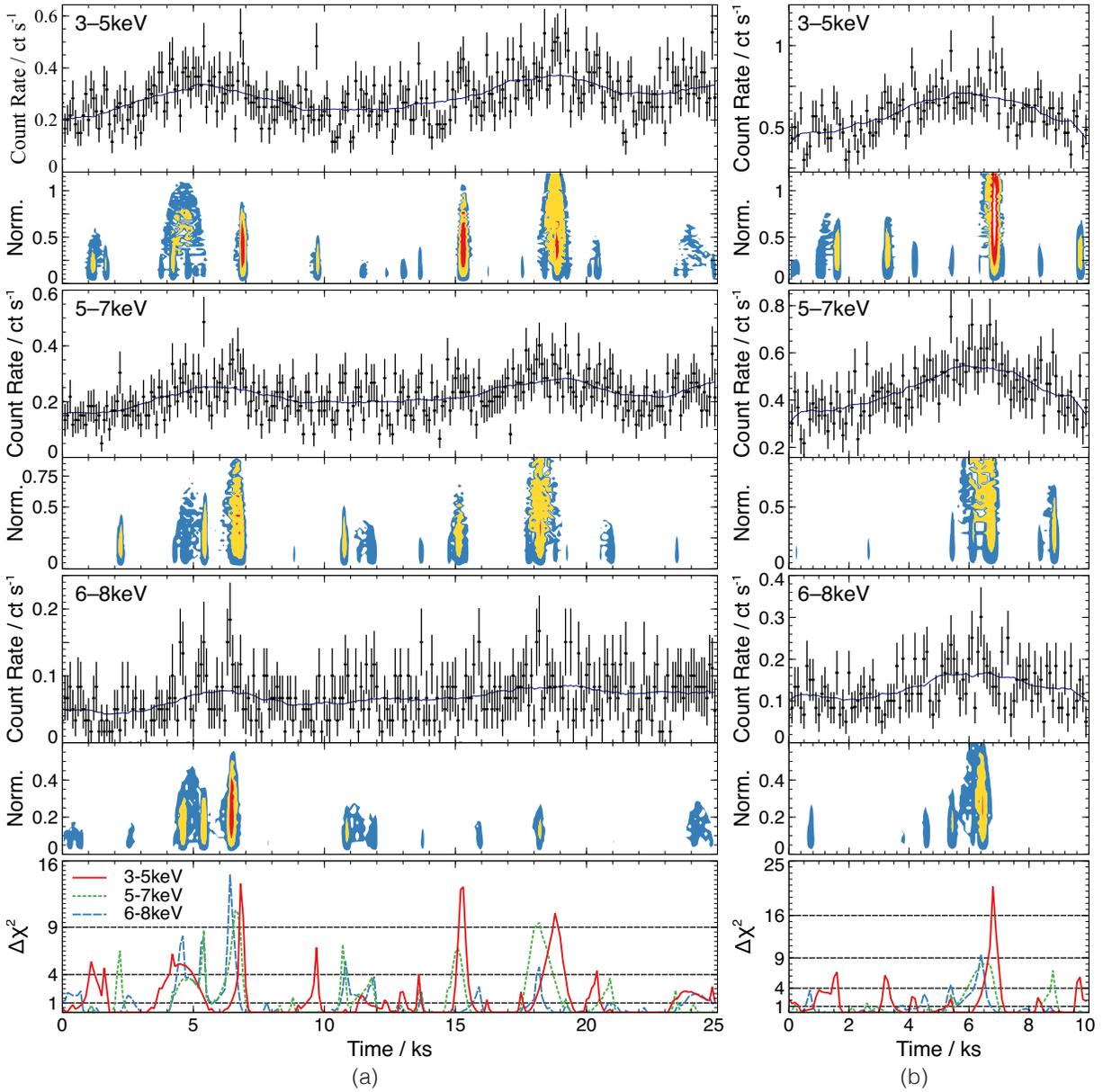

Extended Data Figure 3: Detection of short peaks in the X-ray light curves. (a) The raw light curves obtained from the *XMM-Newton* EPIC pn camera in the 3-5 keV, 5-7 keV and 6-8 keV energy bands, in 100 s time bins. The blue line shows the smoothed light curve model, with respect to which the significance of short peaks is assessed. The panel beneath each light curve shows the reduction in the $\chi^2$ statistic over this model, when a short Gaussian peak is added in different time bins with different normalization. Shaded regions correspond to 1, 2 and 3σ detections. The bottom panel shows the maximum reduction in $\chi^2$ when a flare is added in each time bin. The peaks of interest that are offset between the bands appear at 6.8 ks and 18.8 ks. (b) The same, with light curves in each energy band summed between the first and second flare, aligning the 3-5 keV peaks at 6.8 ks and 18.8 ks.

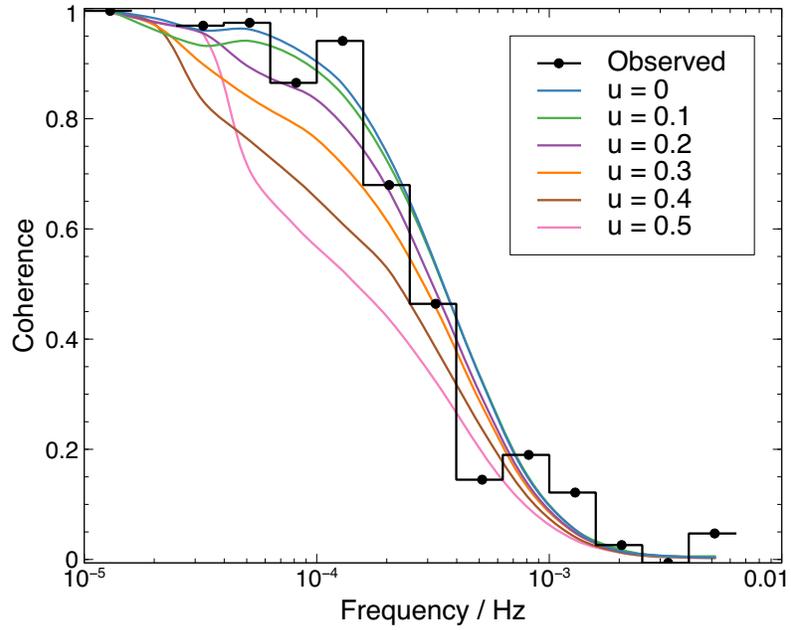

**Extended Data Figure 4: The coherence between the 3-5 keV and 1-3 keV light curves.** The coherence describes the fraction of the variability in each light curve, at different Fourier frequencies, that can be predicted based upon a linear transformation (for example the reverberation response) of the variability in the other light curve. The observed coherence is compared with predictions of pairs of random, red-noise light curve between which a fraction u of variability is uncorrelated. Much of the drop in coherence at high frequencies is due to Poisson noise and the high values of the coherence measured at low frequencies set tight constraints on the level of uncorrelated variability that can